\documentstyle[psfig,prd,aps]{revtex}
\begin{document}
\draft

\title{Numerical simulations of Gowdy spacetimes on ${S^2} \times {S^1}
\times R$}

\author{David Garfinkle
\thanks {Email: garfinkl@oakland.edu}}
\address{
\centerline{Department of Physics, Oakland University,
Rochester, Michigan 48309}}

\maketitle

\null\vspace{-1.75mm}

\begin{abstract}

Numerical simulations are performed of the approach to the singularity
in Gowdy spacetimes on ${S^2} \times {S^1} \times R$.  The behavior
is similar to that of Gowdy spacetimes on ${T^3} \times R$.  In
particular, the singularity is asymptotically velocity term dominated,
except at isolated points where spiky features develop.  
\end{abstract}


\section{Introduction}

There have been several numerical investigations of the approach to the
singularity in inhomogeneous 
cosmologies.
\cite{berger,berger2,bg,weaveretal,berger3,berger4,allofus,stewart}
In general, it is
found that (except at isolated points) the approach to the singularity
is either asymptotically velocity term dominated\cite{im} 
(AVTD) or is oscillatory.
In the oscillatory case, there are epochs of velocity term dominance
punctuated by short ``bounces.''  
The most extensively studied 
inhomogeneous cosmology
is the Gowdy spacetime\cite{gowdya,gowdy} on ${T^3} \times R$.  
Here the approach
to the singularity is AVTD except at isolated points.  The Gowdy spacetimes
on ${T^3} \times R$ are especially well suited to a numerical treatment for
the following reasons: (i) Due to the presence of two Killing fields, the 
metric components depend on only two spacetime coordinates. (ii) The 
constraint equations are easy to implement.  (iii) The boundary conditions
are particularly simple, just periodic boundary conditions in the one
nontrivial spatial direction.

The original work of Gowdy\cite{gowdy} treated spatially compact spacetimes
with a two parameter spacelike isometry group.  Gowdy showed that, for these
spacetimes, the topology of space must be $T^3$ or $S^3$ or ${S^2} \times
{S^1}$.  Given the numerical results for the $T^3$ case, it is natural to
ask what happens in the other two cases.  In a recent paper\cite{oandr}
Obregon and Ryan note that the Kerr metric between the outer and inner
horizons is a Gowdy spacetime with spatial topology ${S^2} \times {S^1}$.
They analyze the behavior of this spacetime and speculate that there
may be significant differences between the behavior of Gowdy spacetimes
on ${T^3} \times R$ and on ${S^2} \times {S^1} \times R$.

A numerical simulation of the ${S^2} \times {S^1}$ case presents some 
difficulties that are absent in the $T^3$ case.  
The constraint equations
become more complicated, and there are difficulties associated with boundary
conditions.  In the $T^3$ case, the Killing fields are nowhere vanishing.
However, in the ${S^2} \times {S^1}$ case, one of the Killing fields
vanishes at the north and south poles of the $S^2$.  Smoothness of the
metric at these axis points then requires that the metric components 
behave in a particular way at these points.  A computer code to evolve
the ${S^2} \times {S^1}$ case therefore must enforce these smoothness
conditions as boundary conditions, and must do so in such a way that the
evolution is both stable and accurate.  These issues are similar to those
encountered in the numerical evolution of axisymmetric spacetimes, and 
the techniques presented here for Gowdy spacetimes should be useful for
axisymmetric spacetimes as well.

This paper presents the results of numerical simulations of Gowdy spacetimes
on ${S^2} \times {S^1} \times R$.  Section 2 presents the metric and 
vacuum Einstein equations in a form suitable for numerical evolution.
The numerical technique is presented in section 3, with the results given
in section 4.

\section{Metric and Field Equations}

The Gowdy metric on ${S^2} \times {S^1} \times R$ has the form\cite{gowdy}
\begin{equation}
\label{metric}
d {s^2} = {e^M} \; ( - \, d {t^2} \, + \, d {\theta ^2} ) \; + \; 
\sin t \; \sin \theta \; \left ( {e^L} {{[ d \phi \, + \, Q \,
d \delta ]}^2} \; + \; {e^{- L}} d {\delta ^2} \right )  \; \; \; .
\end{equation}
Here the metric functions $M, \, L $ and $Q$ depend only on
$t$ and $\theta$.  Thus our two Killing fields are $(\partial /\partial
\phi )$ and $(\partial /\partial \delta )$.  The coordinates $\phi $
and $\delta $ are identified with period $2 \pi $, with $\delta $ the
coordinate on the $S^1$ and $(\theta , \phi )$ the coordinates on the $S^2$.

Here the ``axis'' points are at $\theta = 0$ and $\theta = \pi $.  The
spacetime singularities (``big bang'' and ``big crunch'') are at $t=0$
and $t=\pi $.  This form of the metric presents difficulties for a 
numerical treatment.  Smoothness at the axis requires divergent behavior
in the functions $L$ and $M$.  Furthermore, the spactimes 
singularities occur at finite values of the time coordinate.  This is likely
to lead to bad behavior of the numerical simulation near $t=0$ or
$t=\pi$.  These difficulties are overcome with a new choice of metric
functions and time coordinate.  Define the new metric functions 
$P$ and $\gamma $ by
\begin{equation}
\label{pdef}
P \equiv L \; - \; \ln \sin \theta \; \; \; ,
\end{equation}
\begin{equation}
\label{gamdef}
2 \gamma \equiv M \; - \; ( P \, + \, \ln \sin t ) \; \; \; .
\end{equation}
Define the new time coordinate $\tau $ by
\begin{equation}
\label{taudef}
\tau \equiv - \, \ln \tan (t/2) \; \; \; .
\end{equation}
The metric then takes the form
\begin{equation}
\label{newmetric}
d {s^2} = {1 \over {\cosh \tau }} \; \left \{ {e^P} \left [
{e^{2 \gamma }} \; \left ( {{- \, d {\tau ^2}} \over {{\cosh ^2} \tau}} \; 
+ \; d {\theta ^2} \right ) \; + \; {\sin ^2} \theta \, {{(d \phi \, + \,
Q \, d \delta )}^2}\right ] \; + \; {e^{- P}} \; d {\delta ^2} \right \}
\; \; \; .
\end{equation}

Smoothness of the metric at the axis is equivalent to the
requirement that $P, \, Q$ and
$\gamma $ be smooth functions of $\cos \theta $ with $\gamma $ vanishing
at $\theta = 0 $ and $\theta = \pi $.  Note that for $f$ any smooth
function of $\cos \theta $, it follows that $df/d \theta = 0 $ at 
$\theta = 0 $ and $\theta = \pi$.  

As in the $T^3$ case, the vacuum Einstein field equations 
become evolution
equations for $P$ and $Q$ and ``constraint'' equations that determine
$\gamma $.  The evolution equations are
\begin{equation}
\label{evolveP}
{P_{\tau \tau}} = {e^{2 P}} \; {\sin ^2} \theta {{({Q_\tau})}^2} \; + \;
{1 \over {{\cosh ^2} \tau}} \; \left [ {P_{\theta \theta }} \; + \; 
\cot \theta \, {P_\theta } \; - \; 1 \; - \; {e^{2 P}} \, {\sin ^2} \theta
\, {{({Q_\theta })}^2} \right ] \; \; \; ,
\end{equation}
\begin{equation}
\label{evolveQ}
{Q_{\tau \tau }} = - \, 2 {P_\tau} \, {Q_\tau } \; + \; {1 \over {{\cosh ^2}
\tau }} \; \left [ {Q_{\theta \theta }} \; + \; 3 \, \cot \theta \,
{Q_\theta} \; + \; 2 \, {P_\theta } {Q_\theta } \right ] \; \; \; .
\end{equation}
Here a subscript denotes partial derivative with respect to the corresponding
coordinate.  Note that, as in the $T^3$ case, the evolution equations have
no dependence on $\gamma$.  

The constraint equations are 
\begin{equation}
\label{Aconstr}
\cot \theta \, {\gamma _\tau } \; - \; \tanh \tau \, {\gamma _\theta} = A
\; \; \; ,
\end{equation}
\begin{equation}
\label{Bconstr}
{{\cot \theta } \over {{\cosh ^2} \tau }} \; {\gamma _\theta }
\; - \; \tanh \tau \, {\gamma _\tau } = B  \; \; \; ,
\end{equation}
where the quantities $A$ and $B$ are given by
\begin{equation}
\label{Adef}
2 A \equiv \tanh \tau \, {P_\theta } \; + \; {P_\tau } \, {P_\theta } \; + \;
{e^{2 P}} \, {\sin ^2} \theta \, {Q_\tau } \, {Q_\theta } \; \; \; ,
\end{equation}
\begin{equation}
\label{Bdef}
4 B \equiv 2 \, \tanh \tau \, 
{P_\tau } \; + \; {{({P_\tau })}^2} \; + \; {e^{2P}} \, {\sin ^2} \theta
\, {{({Q_\tau })}^2} \; + \; {\tanh ^2} \tau \; - \; 4
+ \; {1 \over {{\cosh ^2} \tau }} \; \left [ {{({P_\theta })}^2} \; + \;
{e^{2P}} \, {\sin ^2} \theta \, {{({Q_\theta })}^2} \right ] \; \; \; .
\end{equation}
Solving equations (\ref{Aconstr}) and (\ref{Bconstr}) 
for $\gamma _\theta $ and $\gamma _\tau$ we find
\begin{equation}
\label{gamth}
{\gamma _\theta } = {{{\cosh ^2} \tau \, (A \, \tanh \tau \, + \, B \,
\cot \theta )} \over {{\cot^2} \theta \, - \, {\sinh ^2} \tau }} \; \; \; ,
\end{equation}
\begin{equation}
\label{gamtau}
{\gamma _\tau} = {{A \, \cot \theta \, + \, B \, \sinh \tau \, \cosh \tau }
\over {{\cot ^2} \theta \, - \, {\sinh ^2} \tau }} \; \; \; .
\end{equation}
Given a solution of the evolution equations (\ref{evolveP}) and  
(\ref{evolveQ}) for $P$ and $Q$, equations (\ref{gamth})
and (\ref{gamtau}) 
and the smoothness condition that $\gamma = 0 $ at $\theta = 0$ completely
determine $\gamma$.  Actually, equations (\ref{gamth}) and
(\ref{gamtau}) seem to be in danger of
over determining $\gamma $, but the integrability condition for these
equations is automatically satisfied as a consequence of the evolution
equations for $P$ and $Q$.  There are, however, two remaining difficulties
with the equations for $\gamma$.  The first has to do with the fact that
the denominator in equations (\ref{gamth}) and (\ref{gamtau})
vanishes when $|\cot \theta | = \sinh \tau $.  Smoothness of the metric
then requires that the numerators of these equations vanish whenever the
denominator does.  This places conditions on $P$ and $Q$.  If these 
conditions are satisfied for the initial data, the evolution equations
will preserve them.  The second difficulty has to do with the fact that
$\gamma $ must vanish at $\theta = \pi $ as well as $\theta = 0 $.  
Integrating equation (\ref{gamth}) from $0$ to $\pi$ it then
follows that we must have
\begin{equation}
\label{integral}
{\int _0 ^\pi} \; {{{\cosh ^2} \tau \, (A \, \tanh \tau \, + \, B \,
\cot \theta ) \; d \theta} \over {{\cot^2} \theta \, - \, {\sinh ^2} \tau }}
\; = 0 \; \; \; .
\end{equation}
If this condition is satisfied by the initial data, then the evolution
equations will preserve it.

In summary, the initial data for $P$ and $Q$ are not completely freely
specifiable.  They must satisfy conditions at the points where 
$|\cot \theta | = \sinh \tau $ as well as an integral condition.  
Given initial data satisfying these conditions, the evolution equations
(\ref{evolveP}) and (\ref{evolveQ})
then determine $P$ and $Q$ and the constraint equations 
(\ref{gamth}) and (\ref{gamtau}) then determine
$\gamma $.

\section{Numerical Methods}

We now turn to the numerical methods used to implement the evolution
equations.  We begin by casting the equations in first order form
by introducing the quantities $V \equiv {P_\tau }$ and $W \equiv {Q_\tau}$.
These quantities then satisfy the equations
\begin{equation}
\label{evolveV}
{V_\tau} = {e^{2 P}} \; {\sin ^2} \theta {W^2} \; + \;
{1 \over {{\cosh ^2} \tau}} \; \left [ {P_{\theta \theta }} \; + \; 
\cot \theta \, {P_\theta } \; - \; 1 \; - \; {e^{2 P}} \, {\sin ^2} \theta
\, {{({Q_\theta })}^2} \right ] \; \; \; ,
\end{equation}
\begin{equation}
\label{evolveW}
{W_\tau } = - \, 2 \, V \, W \; + \; {1 \over {{\cosh ^2}
\tau }} \; \left [ {Q_{\theta \theta }} \; + \; 3 \, \cot \theta \,
{Q_\theta} \; + \; 2 \, {P_\theta } {Q_\theta } \right ] \; \; \; .
\end{equation}

Thus the evolution equations have the form $ {{\vec X}_\tau} = {\vec F}
( {\vec X},\tau )$.  We implement these equations using an iterative
Crank-Nicholson scheme.  Given $\vec X$ at time $\tau$, we define
$ {{\vec X}_0} (\tau + \Delta \tau ) \equiv {\vec X} (\tau )$ and then
iterate the equation
\begin{equation}
\label{CNdef}
{{\vec X}_{n+1}} (\tau + \Delta \tau ) = {\vec X} (\tau ) \; + \; 
{{\Delta \tau} \over 2} \; \left [ {\vec F} ({\vec X}(\tau ),\tau ) \; + \;
{\vec F} ( {{\vec X}_n}(\tau + \Delta \tau ),\tau + \Delta \tau) 
\right ] \; \; \; .
\end{equation}
In principle, one should iterate until some sort of convergence is 
achieved.  In practice, we simply iterate 10 times.  We use 
$\Delta \tau = \Delta \theta / 2 $ where $\Delta \theta $ is the
spatial grid spacing.

The spatial grid is as follows: Let $n_\theta $ be the number of
spatial grid points.  Then we choose $\Delta \theta = \pi / ({n_\theta}
- 2 )$ and ${\theta _i} = (i - 1.5) \Delta \theta $.  Thus in addition
to the ``physical zones'' for $i = 2, 3, \dots , {n_\theta } - 1 $, we 
have two ``ghost zones'' at ${\theta _1} = - \, \Delta \theta / 2$ and
${\theta _{n_\theta }} = \pi + (\Delta \theta /2)$.  The ghost zones
are not part of the spacetime: variables there are set by boundary
conditions.  For any quantity $S$, define ${S_i} \equiv S ({\theta _i})$.
Spatial derivatives are implemented using the usual second order scheme:
\begin{equation}
\label{firstderiv}
{S_\theta } ( {\theta _i}) = {{{S_{i+1}} \, - \, {S_{i-1}}} \over
{2 \Delta \theta }} \; \; \; ,
\end{equation}
\begin{equation}
\label{secondderiv}
{S_{\theta \theta }} ( {\theta _i} ) = {{{S_{i+1}} \, + \, 
{S_{i-1}} \, - \, 2 \, {S_i}} \over {{(\Delta \theta )}^2}} \; \; \; .
\end{equation}

Smoothness of the metric requires that ${P_\theta } = 0 $ at $\theta = 0 $.
Since $\theta = 0 $ is halfway between $i=1$ and $i=2$, we implement this 
condition as ${P_1} = {P_2}$.  Similarly, we use ${Q_1} = {Q_2} $ since
${Q_\theta } = 0 $ at $\theta = 0 $.  Correspondingly, the requirement that
$P_\theta $ and $Q_\theta $ vanish at $\theta = \pi $ is implemented as
${P_{n_\theta }} = {P_{{n_\theta} - 1}} $ and ${Q_{n_\theta }} =
{Q_{{n_\theta} - 1}} $.  These boundary conditions are imposed at each
iteration of the Crank-Nicholson scheme.

\section{Results}

To test the computer code, it is helpful to have some closed form exact
solution of the evolution equations to compare to the numerical evolution
of the corresponding initial data.  In particular, for a second order
accurate evolution scheme, the difference between the numerical solution
and the exact solution should converge to zero as the grid spacing squared.

A polarized Gowdy spacetime is one for which the Killing vectors are 
hypersurface orthogonal.  For our form of the metric, that is
equivalent to the condition $Q=0$.  For polarized Gowdy spacetimes, the
evolution equation (\ref{evolveQ}) for $Q$ is trivially satisfied, 
and the evolution
equation (\ref{evolveP}) for $P$ reduces to the following:
\begin{equation}
\label{evolvePpol}
{P_{\tau \tau}} = {1 \over {{\cosh ^2} \tau}} \; \left [ {P_{\theta \theta }}
\; + \; \cot \theta \, {P_\theta } \; - \; 1 \right ] \; \; \; .
\end{equation}
This is a linear equation which can be solved by separation of variables,
though one must choose only those solutions that satisfy the additional
conditions for smoothness of the metric.

Unfortunately, the polarized solutions do not provide, by themselves, a
very stringent code test: the evolution equation for $Q$ and the
nonlinear terms in the evolution equation for $P$ are not tested at all.
Fortunately, there is a technique, the Ehlers solution generating
technique, which allows us to begin with a polarized solution and 
produce an unpolarized solution.  Let $\bar P$ be any solution of the 
polarized equation (\ref{evolvePpol}) and let $c$ be any constant.  
Define $P$ and $Q$ by
\begin{equation}
\label{Pgen}
P = {\bar P} \; - \; \ln \left [ 1 \; + \; {{\left ( {{c \, {\sin ^2} 
\theta } \over {\cosh \tau }} \; {e^{\bar P}} \right ) }^2} \right ] 
\; \; \; ,
\end{equation}
\begin{equation}
\label{Qtaugen}
{Q_\tau } = {{- \, 2 \, c}\over {{\cosh ^2} \tau }} \; \left ( 
2 \, \cos \theta \; + \; \sin \theta \, {{\bar P}_\theta } \right )
\; \; \; ,
\end{equation}
\begin{equation}
\label{Qthgen}
{Q_\theta } = 2 \, c \, \sin \theta \, \left ( \tanh \tau \; - \; 
{{\bar P}_\tau} \right ) \; \; \; .
\end{equation}
Then $(P, Q)$ is a solution of the unpolarized Gowdy equations 
(\ref{evolveP}) and (\ref{evolveQ}). 

We use the following polarized solution:
\begin{equation}
\label{Ppoltest}
{\bar P} = - \, \ln \cosh \tau \; + \; 2 \, \tau \; \; \; ,
\end{equation}
where $b$ is a constant.  The solution generating technique then yields an
unpolarized solution with $P$ given by equation (\ref{Pgen}) 
and $Q$ given by
\begin{equation}
\label{Qtest}
Q = 4 \, c \, ( 1 \, - \, \tanh \tau ) \, \cos \theta  \; \; \; .
\end{equation}

Figure 1 shows $P$ for the exact solution and the 
numerical evolution.  (Here there are 502 spatial grid points, 
$c = 1$, and the initial data at $\tau = 0$ are evolved to $\tau = 10$.
The results are shown at 51 equally spaced points from $\theta = 0 $ to
$\theta = \pi $).
Figure 2 shows the difference between exact and numerical
solutions.  Here the parameters are as in figure (1), except that
two simulations are run: one with 502 spatial gridpoints and one with 
1002 grid points.  For comparison, the results on the finer grid are
multiplied by a factor of 4.  The results show second order convergence.
(Note: due to the presence of the ghost zones, quantities must be
interpolated on the grids to make a comparison between quantities at
the same values of $\theta$).

We would now like to find the generic behavior of Gowdy spacetimes on
${S^2} \times {S^1} \times R$.  In the $T^3$ case\cite{berger} a family
of initial data was chosen and evolved.  It was argued that the behavior
of these spacetimes reflects the generic behavior.  Here, we choose a 
similar family.  The initial data at $\tau = 0 $ are 
$P=0, \, {P_\tau}={v_0} \cos \theta, \, Q=2 \cos \theta , \,
{Q_\tau} = 0$.  Here $v_0$ is a constant.  These data satisfy the
constraint conditions.  Figures 3-8 show the evolution of these data for
various values of the parameter $v_0$.  Here, ${v_0}=2$ in 
figures 3 and 4, 
${v_0}=4$ in figure 5 and 6, and ${v_0}=8$ in figures 
7 and 8.  In all cases, the
range of $\theta $ is $(0,\pi )$, the range of $\tau $ is $(0,10)$ and
the simulation is run with 1002 spatial grid points.  
Note the presence of spiky 
features.  The large $\tau $ behavior of the solutions is the following:
There are functions ${Q_\infty}(\theta)$ and ${v_\infty} (\theta )$
with ${v_\infty} (\theta ) < 1$
such that away from the spiky features we have $Q \to {Q_\infty}(\theta)$
and ${P_\tau} \to {v_\infty}(\theta)$ for large $\tau$.  

The reason for this behavior is not hard to find and is essentially the
same as in the $T^3$ case.  For large $\tau$ and provided that $P$ is
growing no faster than $\tau$, The terms in equations 
(\ref{evolveP}) and (\ref{evolveQ}) proportional to
$1/{\cosh ^2} \tau$ become negligible.  The truncated equations obtained
by neglecting these terms are
\begin{equation}
\label{avtdP}
{P_{\tau \tau}} = {e^{2 P}} \, {\sin ^2} \theta \, {{({Q_\tau})}^2} 
\; \; \; ,
\end{equation}
\begin{equation}
\label{avtdQ}
{Q_{\tau \tau }} = - \, 2 \, {P_\tau} \, {Q_\tau} \; \; \; .
\end{equation}
Equations (\ref{avtdP}) and (\ref{avtdQ}) are 
called the AVTD equations.  They can be solved in 
closed form and have the property that $Q \to {Q_\infty} (\theta )$
and ${P_\tau } \to {v_\infty}(\theta )$ as $\tau \to \infty $.
Solutions of the full equations (\ref{evolveP}) and (\ref{evolveQ}) 
are called AVTD provided that they 
approach solutions of the AVTD equations for large $\tau$.  Thus we
have an explanation of the AVTD behavior provided that we can show
that $P$ grows no faster than $\tau$.  As in the $T^3$ case, if $P$
grows faster than $\tau$ then the term in equation (\ref{evolveP}) 
proportional to
${e^{2P}}/{\cosh ^2} \tau$ will cause a ``bounce'' that leaves $P$
growing less fast than $\tau$.  Thus, an analysis of the large $\tau$
behavior of the evolution equations (\ref{evolveP}) and 
(\ref{evolveQ}), essentially the same as in the 
$T^3$ case, leads to an explanation of the AVTD behavior.

The AVTD behavior can also be explained by an analysis of the local
properties of Gowdy spacetimes on ${S^2} \times {S^1} \times R$.
Define ${S_a} \equiv {\nabla _a} ( \sin t \, \sin \theta )$.  Then
in regions of the spacetime where $S_a$ is timelike, the region is 
locally isometric to a Gowdy spacetime on ${T^3} \times R$.  In regions
where $S_a$ is spacelike, the region is locally isometric to a 
cylindrical wave.  Thus, the behavior that we should expect in the 
${S^2} \times {S^1}$ case is a combination of the  the behavior of the
$T^3$ case and the behavior of cylindrical waves.  Furthermore, for 
any point on the $S^2$ except the poles, as the singularity is approached,
$S_a$ becomes timelike at that point.  Thus the asymptotic behavior as
the singularity is approached in the ${S^2} \times {S^1}$ case should  be
the same as in the $T^3$ case.

We now turn to an analysis of the spiky features seen in the metric 
functions $P$ and $Q$.  The argument of the previous paragraph indicates
that these features are essentially the same as those seen in the $T^3$ 
case.  In fact, these features can be explained using the evolution equations
(\ref{evolveP}) and (\ref{evolveQ})
as was done in the $T^3$ case.\cite{bg}  For large $\tau $, 
it follows from equation (\ref{evolveQ}) that 
$ {Q_\tau } \approx {\Pi _Q} (\theta ) \, {e^{ - 2 P }} $ for some function
$ {\Pi _Q} (\theta ) $.  Then, using this result in equation 
(\ref{evolveP}) we have an approximate evolution 
equation for $P$:
\begin{equation}
\label{BGapprox}
{P_{\tau \tau }} \approx {\sin ^2} \theta \; \left [ {e^{ - 2 P}} \,
{{({\Pi _Q})}^2} \; - \; {{e^{2 P}} \over {{\cosh ^2} \tau }} \; 
{{({Q_\theta })}^2} \right ] \; \; \; .
\end{equation}
These terms eventually drive $P_\tau$ to the range between $0$ and $1$.
However, at a point $\theta _1$ where $Q_\theta $ vanishes, $P_\tau $
can be greater than $1$.  This leads to a spiky feature in $P$, since
${P_\tau } > 1 $ at $\theta _1 $ but ${P_\tau } < 1 $ at points near
$\theta _1$.  This sort of spiky feature is illustrated in figure 9.
Correspondingly, at a point $\theta _2$ where $\Pi _Q$ vanishes,
$P_\tau $ can be less than zero.  This leads to sharp features in 
$P$ since ${P_\tau} > 0 $ at points near $\theta _2$.  Also since the
region where $P < 0 $ leads to rapid growth in $Q$, there is a sharp
feature in $Q$.  This sort of feature is illustrated in figure 10.

In summary, a numerical treatment of Gowdy spacetimes on ${S^2} \times
{S^1} \times R$ reveals that they are very similar to Gowdy spacetimes
on ${T^3} \times R$.  In particular, they show the same behavior of 
AVTD behavior almost everywhere, and they have the same sort of spiky
features at isolated points.

\section{Acknowledgements}

I would like to thank Beverly Berger, Vince Moncrief and G. Comer Duncan for
helpful discussions. I would also like to thank the Institute for
Theoretical Physics at Santa Barbara for hospitality.  This work was
partially supported by NSF grant PHY-9722039 to Oakland University.

\begin{figure}[bth]
\begin{center}
\makebox[4in]{\psfig{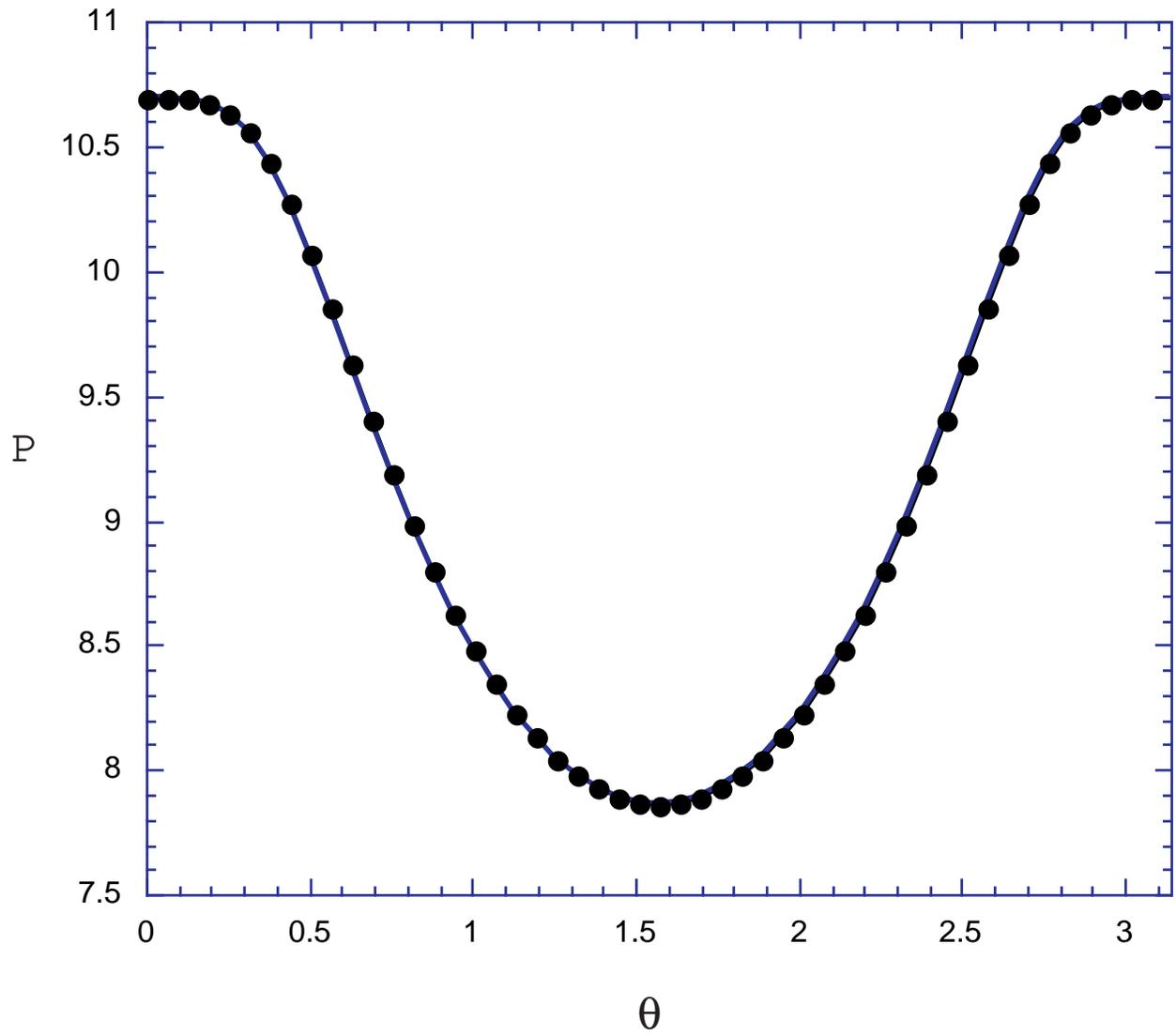}}
\caption{The analytic values (line) and numerical values (dots) of $P$
are plotted vs. $\theta $ at $\tau = 10$.}
\label{fig1}
\end{center}
\end{figure}
\vfill\eject 

\begin{figure}[bth]
\begin{center}
\makebox[4in]{\psfig{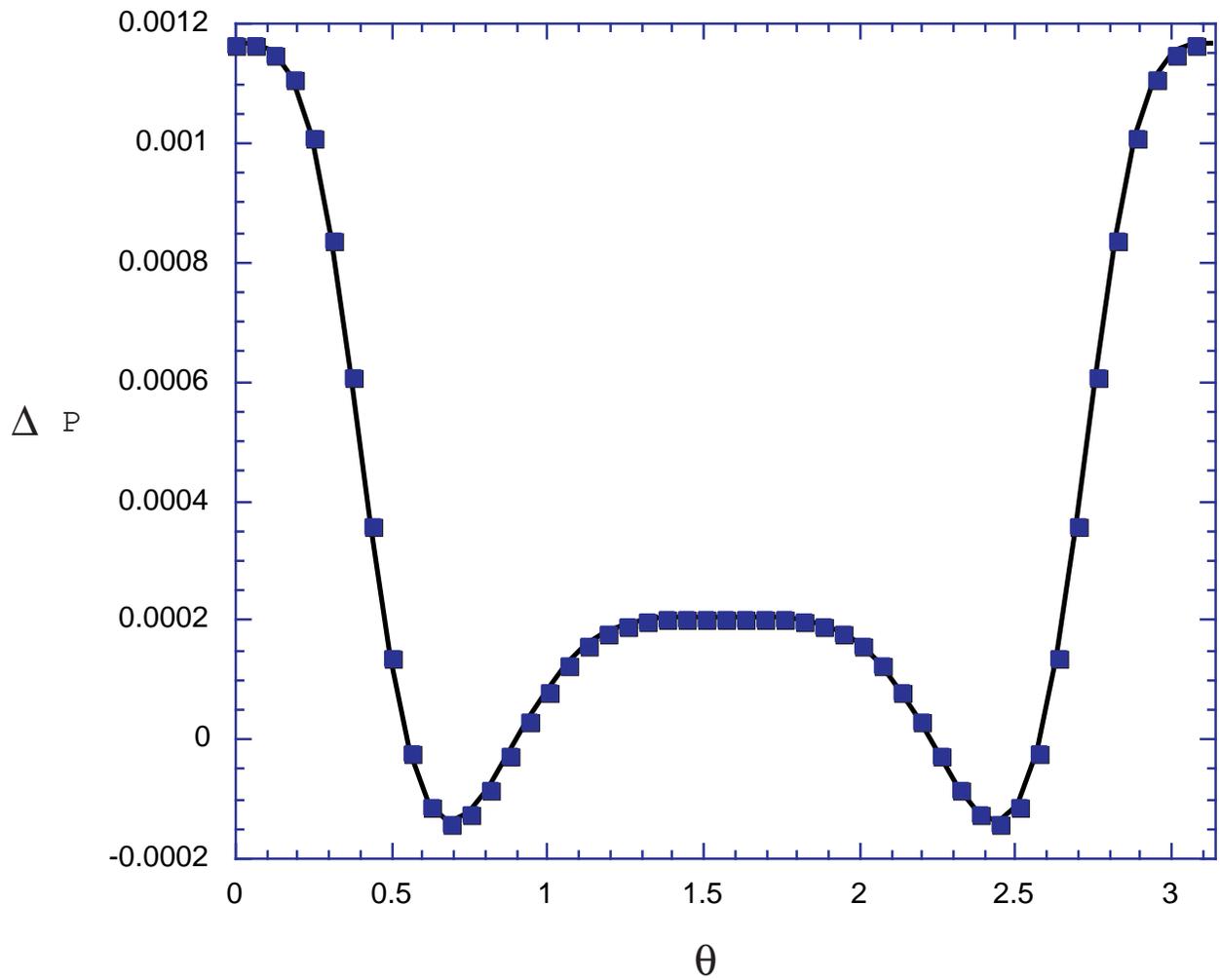}}
\caption{The differences between analytic and numerical values for $P$
are compared for two different resolutions.  The line is $\Delta P$ for
502 grid points and the squares are 4 times $\Delta P$ for 1002 grid
points.  The fact that these curves agree shows second order
convergence}
\label{fig2}
\end{center}
\end{figure}
\vfill\eject

\begin{figure}[bth]
\begin{center}
\makebox[4in]{\psfig{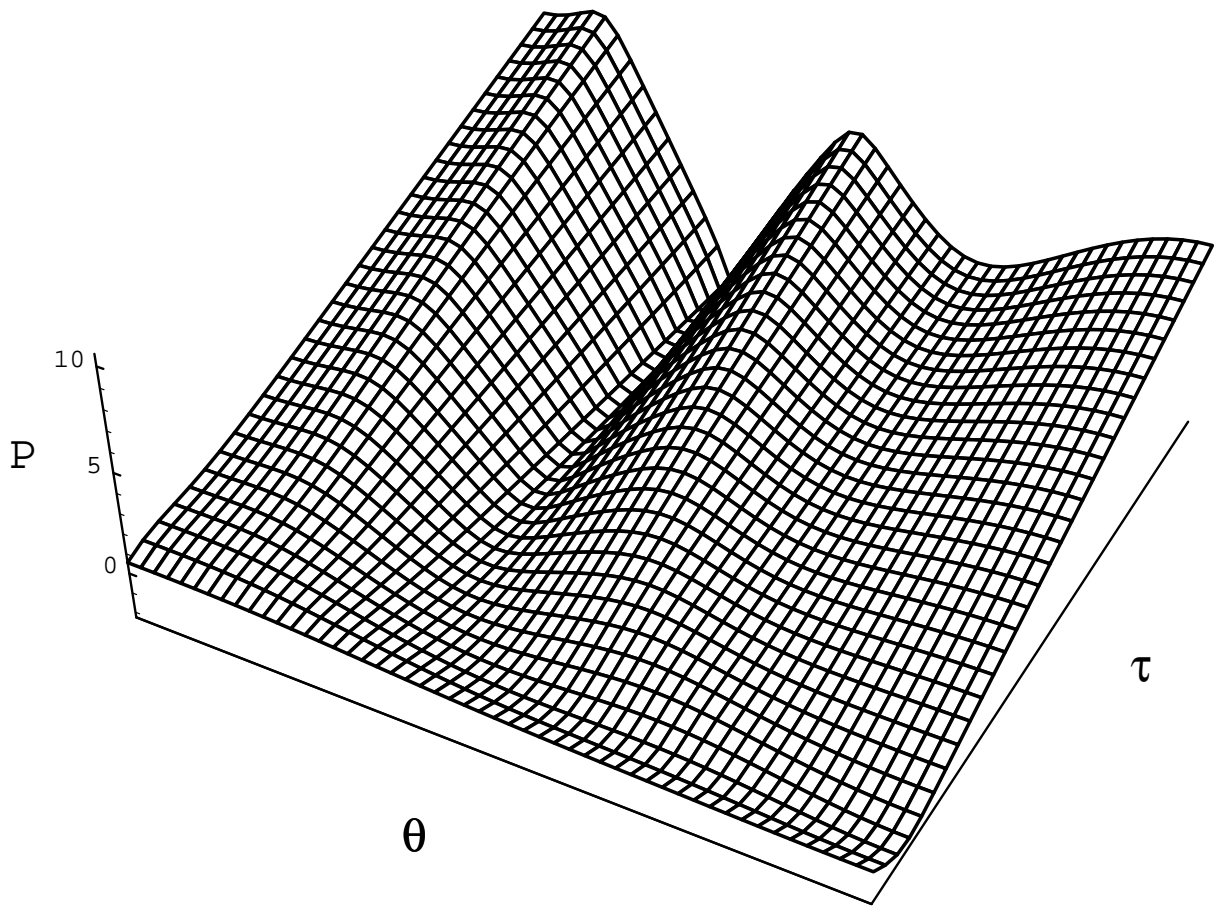}}
\caption{The evolution of $P$ for initial data
with ${v_0} = 2$.}
\label{fig3}
\end{center}
\end{figure}
\vfill\eject

\begin{figure}[bth]
\begin{center}
\makebox[4in]{\psfig{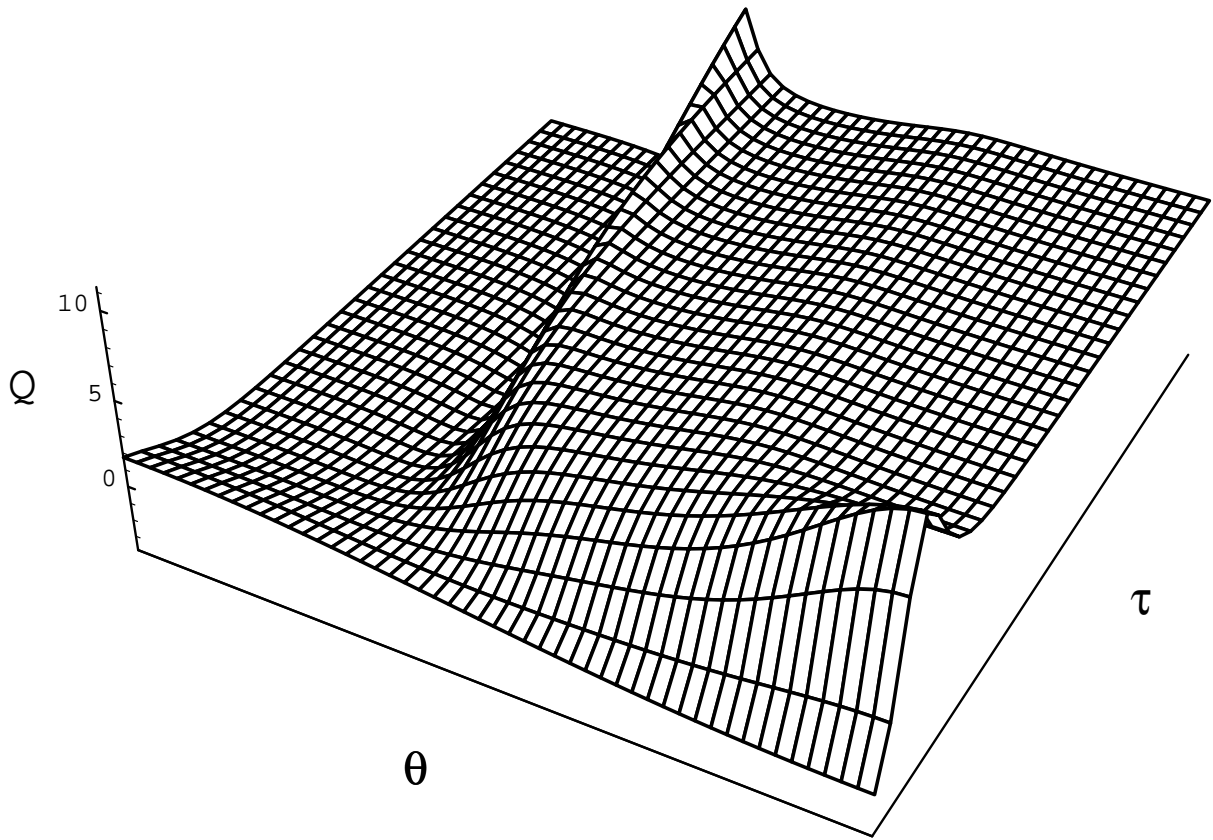}}
\caption{The evolution of $Q$ for initial data
with ${v_0} = 2$.}
\label{fig4}
\end{center}
\end{figure}
\vfill\eject

\begin{figure}[bth]
\begin{center}
\makebox[4in]{\psfig{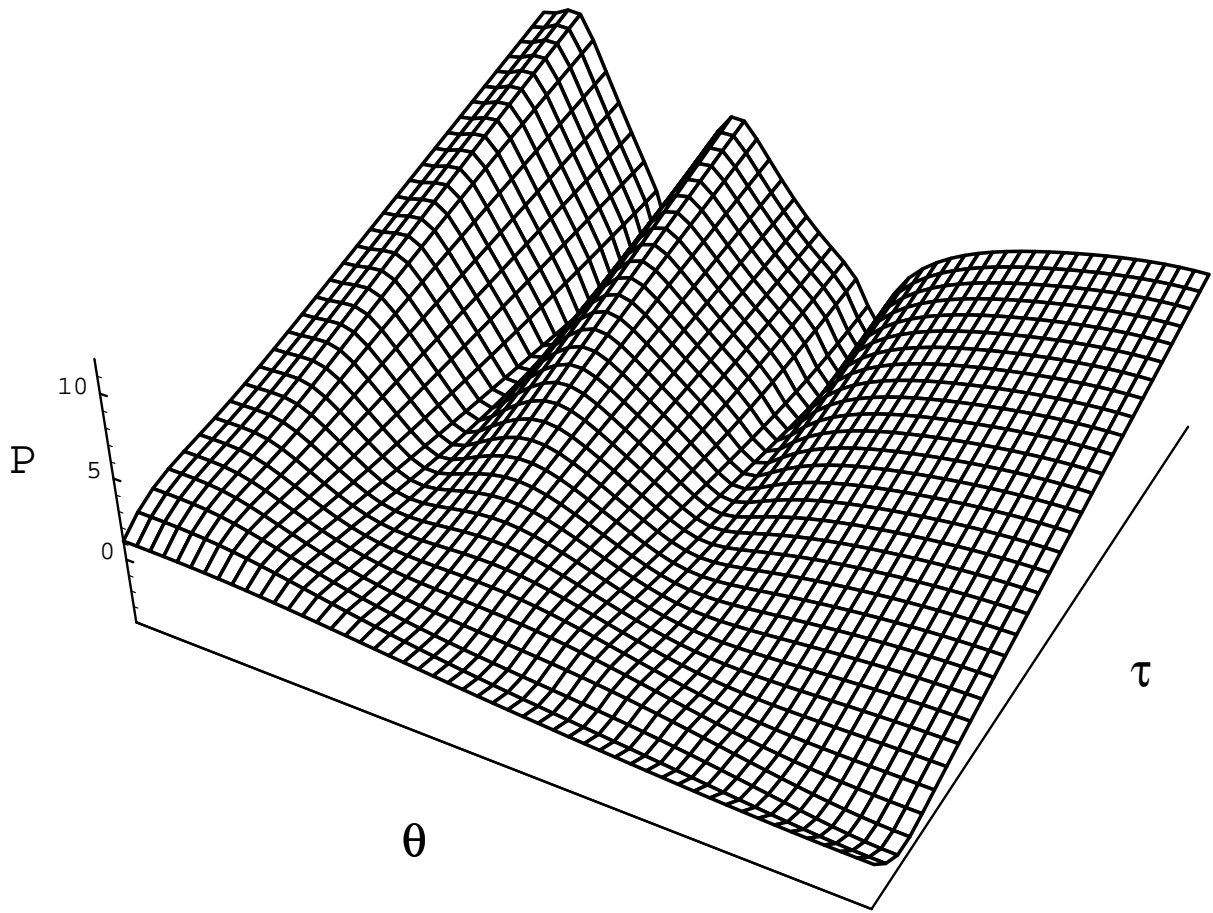}}
\caption{The evolution of $P$ for initial data
with ${v_0} = 4$.}
\label{fig5}
\end{center}
\end{figure}
\vfill\eject

\begin{figure}[bth]
\begin{center}
\makebox[4in]{\psfig{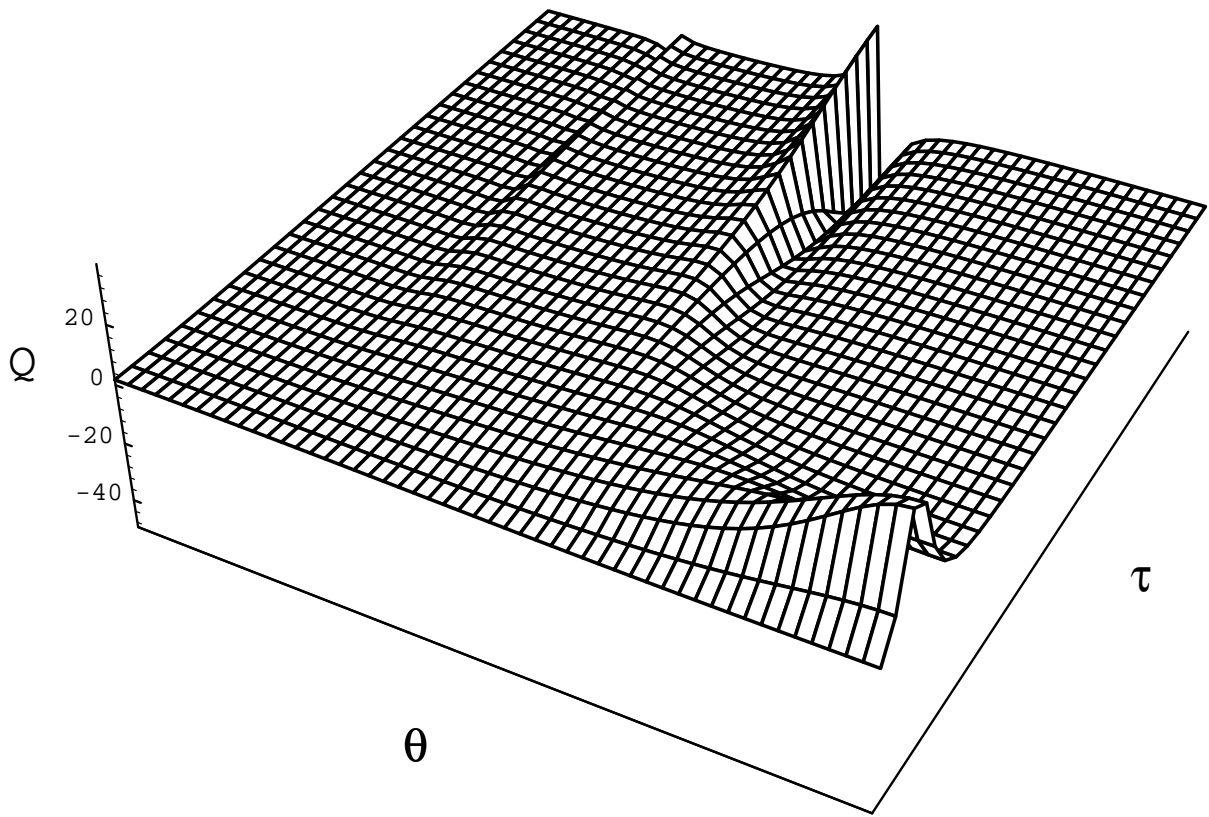}}
\caption{The evolution of $Q$ for initial data
with ${v_0} = 4$.}
\label{fig6}
\end{center}
\end{figure}
\vfill\eject

\begin{figure}[bth]
\begin{center}
\makebox[4in]{\psfig{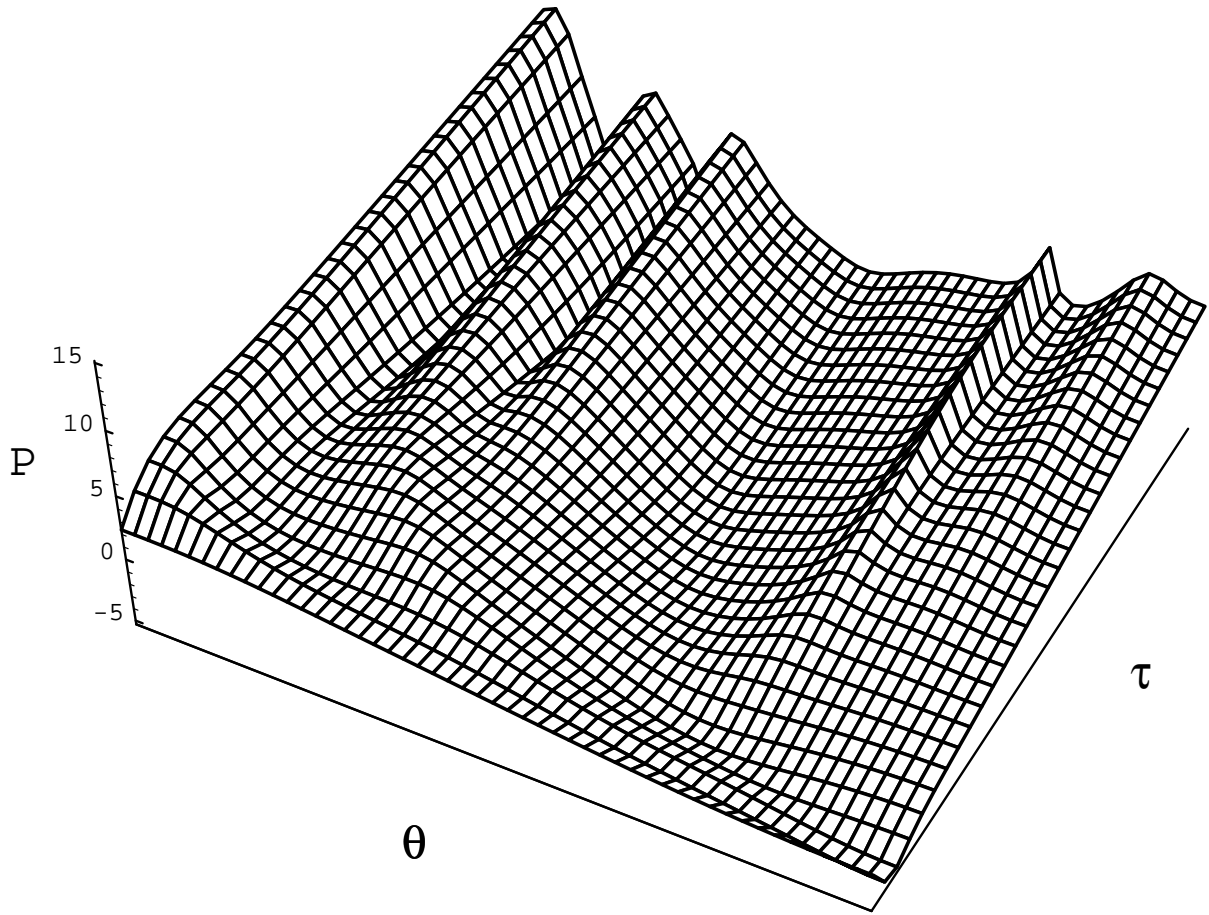}}
\caption{The evolution of $P$ for initial data
with ${v_0} = 8$.}
\label{fig7}
\end{center}
\end{figure}
\vfill\eject

\begin{figure}[bth]
\begin{center}
\makebox[4in]{\psfig{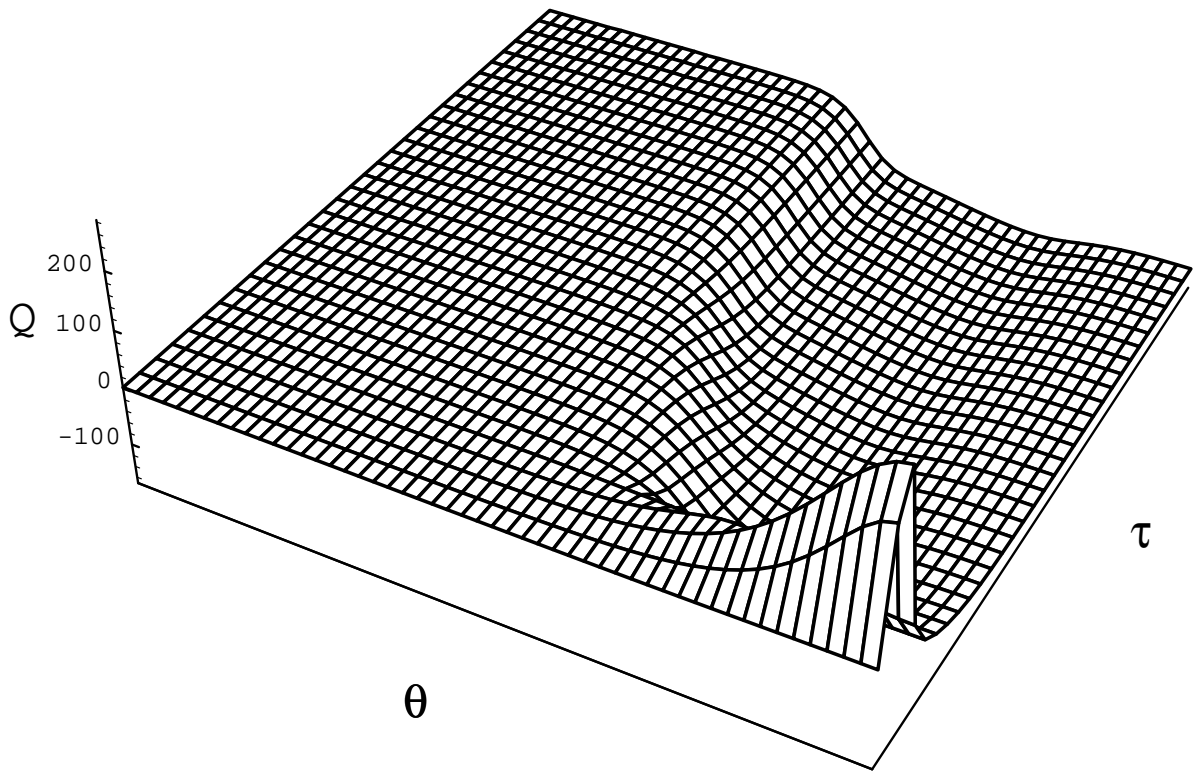}}
\caption{The evolution of $Q$ for initial data
with ${v_0} = 8$.}
\label{fig8}
\end{center}
\end{figure}
\vfill\eject

\begin{figure}[bth]
\begin{center}
\makebox[4in]{\psfig{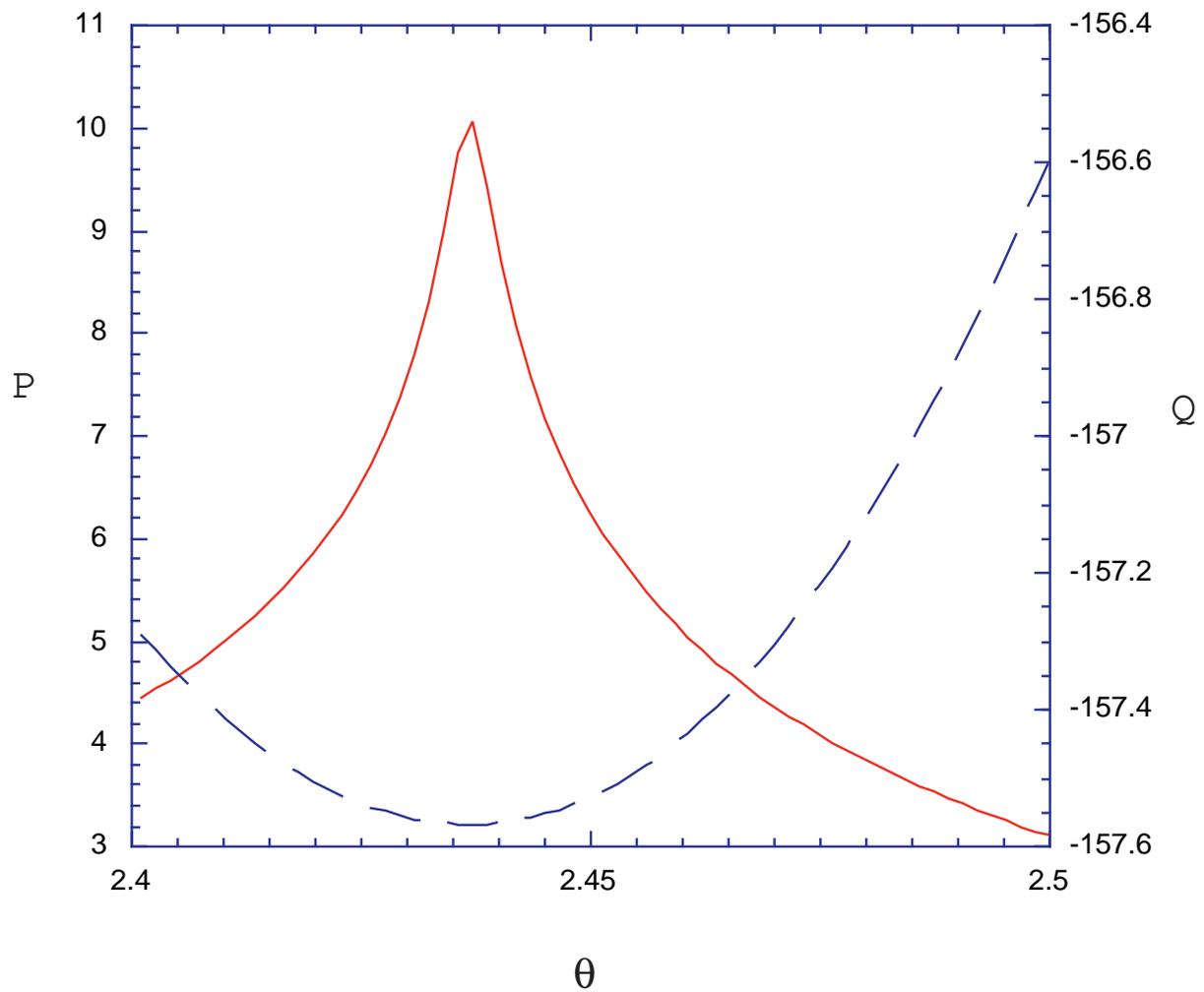}}
\caption{a spiky feature in $P$ (solid line) occurs where $Q$
(dashed line) has an extremum.
Here, ${v_0}=8, \, \tau = 10$ and the simulation is run with 2002
spatial grid points.} 
\label{fig9}
\end{center}
\end{figure}
\vfill\eject

\begin{figure}[bth]
\begin{center}
\makebox[4in]{\psfig{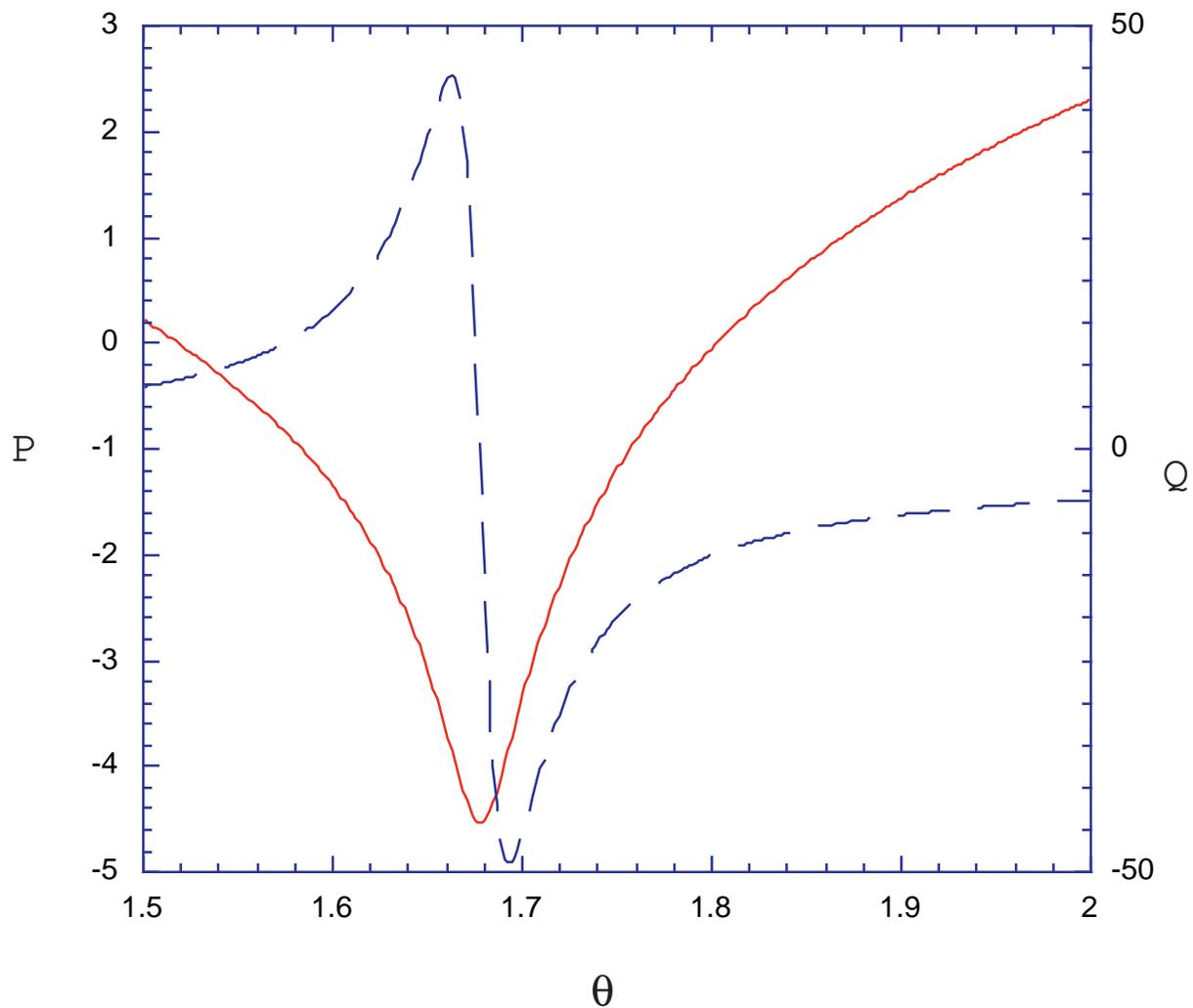}}
\caption{a sharp feature in $Q$ (dashed line) occurs at a downward
spike in $P$ (solid line).  Here, ${v_0}=4, \, \tau = 10$ and the
simulation is run with 2002 spatial grid points.}
\label{fig10}
\end{center}
\end{figure}
\vfill\eject


\begin{references}

\bibitem{berger}
B.K. Berger and V. Moncrief, Phys. Rev. {\bf D48}, 4676  (1993)

\bibitem{berger2}
B.K. Berger, D. Garfinkle and V. Swamy, Gen. Rel. Grav. {\bf 27},
511 (1995)

\bibitem{bg}
B.K. Berger and D. Garfinkle, Phys. Rev. {\bf D57}, 4767 (1998)

\bibitem{weaveretal}
M. Weaver, J. Isenberg and B.K. Berger, Phys. Rev. Lett. 
{\bf 80}, 2984 (1998)

\bibitem{berger3}
B.K. Berger and V. Moncrief, Phys. Rev. {\bf D57}, 7235 (1998)

\bibitem{berger4}
B.K. Berger and V. Moncrief, Phys. Rev. {\bf D58}, 064023 (1998)

\bibitem{allofus}
B.K. Berger, D. Garfinkle, J. Isenberg, V. Moncrief and M. Weaver,
Mod. Phys. Lett. {\bf A13}, 1565 (1998)

\bibitem{stewart}
S.D. Hern and J.M. Stewart, Class. Quantum Grav. {\bf 15}, 1581 (1998) 

\bibitem{im}
For a precise definition of AVTD as well as a proof that polarized
Gowdy spacetimes are AVTD see
J. Iseberg and V. Moncrief, Ann. Phys. (N.Y.) {\bf 199}, 84 (1990) 

\bibitem{gowdya}
R.H. Gowdy, Phys. Rev. Lett. {\bf 27}, 826 (1971)

\bibitem{gowdy}
R.H. Gowdy, Ann. Phys. {\bf 83}, 203 (1974)

\bibitem{oandr}
O. Obregon and M.P. Ryan, gr-qc/9810068


\end{references}
\end{document}